\documentclass[runningheads]{llncs}
\usepackage[T1]{fontenc}
\usepackage{xcolor}
\usepackage{graphicx}
\usepackage{subcaption}
\usepackage{booktabs}
\begin{document}
\title{Interplay between Emotional Dynamics and Network Structure on the Social Media Vent}
\titlerunning{Emotional Dynamics and Network Structure on Vent}
%
\author{Yuina Takahashi\inst{1}\and
Sho Tsugawa\inst{2}}
\authorrunning{Y. Takahashi et al.}
%
\institute{ 
Graduate School of Science and Technology, University of Tsukuba \\
1-1-1 Tennodai, Tsukuba, Ibaraki 305-8573, Japan \\
\email{y.takahashi@snlab.cs.tsukuba.ac.jp}\\
\and
Institute of Systems and Information Engineering, University of Tsukuba\\
1-1-1 Tennodai, Tsukuba, Ibaraki 305-8573, Japan \\
\email{s-tugawa@cs.tsukuba.ac.jp}}
\maketitle              
\begin{abstract}
The sharing and contagion of emotions on social media influence online interactions and users' psychological states. This study analyzes emotional dynamics on Vent, an emotion-sharing social media platform where users explicitly assign emotional labels to their own posts. Using a large-scale dataset, we examine how emotions in users' pre-posting timelines are associated with their subsequent emotional labels and how such associations vary across users. Our analysis reveals three key findings. First, users' subsequent emotional labels are associated with the emotional composition of their pre-posting timelines, with same-category emotions being overrepresented before posts in all analyzed categories. Second, several cross-category associations are observed; for example, Surprise was overrepresented before Fear posts; Affection and Happiness were overrepresented before Anger posts; and Affection was overrepresented before Sadness posts. Third, users differ in their degree of alignment with timeline emotional fluctuations, and highly aligned users tend to be located close to one another in the network. These findings provide large-scale observational evidence that emotional expression on Vent is associated with both short-term timeline context and network structure.

\keywords{Social Media \and Social Network \and Emotional Contagion \and Vent}
\end{abstract}

\section{Introduction}\label{sec:first}

On social media, interactions among users through the expression and sharing of emotions are highly active, giving rise to phenomena such as emotional homophily \cite{fan2014} and emotional contagion \cite{ferrara2015,coviello2014,kramer2014,fan2016,xing2025}. 
Emotional homophily is defined as the tendency for individuals with similar emotional dispositions to connect with one another, while emotional contagion refers to the phenomenon where a user's emotional state influences the emotions of others. These phenomena exert a profound impact on community formation and information diffusion within social media ecosystems. 

Extensive research has investigated emotional contagion and emotional dynamics on social media \cite{coviello2014,fan2016,fan2014,ferrara2015,hossain2023,kramer2014,sener2023}. 
Kramer et al. \cite{kramer2014} experimentally showed that emotional content in Facebook News Feeds affects users' subsequent emotional expressions, while Ferrara et al.  \cite{ferrara2015} examined emotional contagion on Twitter by comparing users' posts with the emotional content in their social surroundings. Furthermore, analyses utilizing more granular classifications have revealed that specific emotions, such as anger, tend to spread more rapidly and extensively than others, such as joy \cite{fan2014,fan2016}.

More recent work has examined finer emotional categories: Hossain et al. \cite{hossain2023} analyzed 28 types of emotions in social media comments and proposed a framework for extracting and predicting emotional trends across discussion threads, while Sener et al. \cite{sener2023} analyzed fine-grained emotional dynamics on Twitter using valence, arousal, and dominance.

However, despite these advances, three limitations remain in existing studies of emotional dynamics on social media. First, although some studies have adopted finer emotion categories, much prior work still relies on coarse-grained or binary emotional classifications, such as positive and negative emotions. Second, prior studies have mainly focused on intra-emotional patterns, while cross-emotional associations remain insufficiently explored. Third, individual differences in users' alignment with surrounding emotional fluctuations and their network structure remain unclear. Understanding these patterns may inform healthier social media design.

In this paper, we analyze temporal associations between pre-posting timeline emotions and users' subsequent emotional labels using the {\itshape Vent} dataset \cite{lykousas2019}, which allows for finer emotional granularity than datasets used in many prior studies. In this paper, we address two research questions to examine emotional dynamics on social media: RQ1 investigates how emotions in users' pre-posting timelines are associated with their subsequent emotional labels, and RQ2 examines how users' susceptibility to timeline emotional fluctuations varies and whether highly susceptible users are close to one another in the network.

The contribution of this paper is threefold. First, we analyze temporal associations between pre-posting timeline emotions and users' subsequent emotional labels using a large-scale Vent dataset with explicit emotion labels. Second, we examine both within-category and cross-category emotional associations. Third, we quantify users' alignment with timeline emotional fluctuations and analyze whether highly aligned users are located close to one another in the social network.

\section{The Vent Dataset}

\begin{table}[tb]
\centering
\caption{Target emotional categories, example labels, and the distribution of posts.}
\label{tab:emotional_cat_summary}
\begin{tabular}{llr}
\toprule
Category   & Example Labels                 & Number of Posts (\%) \\ \midrule
Sadness    & Sad, Lonely, Miserable         & 5,122,245 (27.4\%)   \\
Happiness  & Amused, Happy, Excited         & 3,646,174 (19.5\%)   \\
Anger      & Annoyed, Frustrated, Irritated & 2,930,447 (15.7\%)   \\
Fear       & Anxious, Stressed, Afraid      & 2,626,845 (14.1\%)   \\
Affection  & Loving, Needy, Adoring         & 1,770,829 (9.5\%)    \\
Surprise   & Confused, Curious, Surprised   & 1,657,497 (8.9\%)    \\
Positivity & Hyped, Hopeful, Determined     & 927,766 (5.0\%)      \\ \bottomrule
\end{tabular}
\end{table}

Vent, the platform analyzed in this study, is a social media platform where users can assign emotional labels to their own posts. Unlike typical social media platforms, Vent requires users to select a label representing their current emotion, such as Sad or Happy, at the time of posting. This unique feature allows us to analyze subjective emotions explicitly stated by users without the need for sentiment estimation via text analysis.

The public Vent dataset \cite{lykousas2019} used in this study contains data for 934,095 users, 33,623,414 posts, and a social graph consisting of 13,605,522 following edges. It includes 705 types of labels organized into 63 categories, with each post linked to these emotional labels.

From this public dataset, we focused on posts assigned with labels belonging to eight broad categories: Sadness, Fear, Anger, Surprise, Positivity, Affection, Happiness, and Feelings. However, because the Feelings category contains heterogeneous labels such as Tired, Meh, and Sleepy, it was excluded from the category-level analyses. We therefore defined active users as those who posted at least 30 times using labels in the remaining seven categories, and filtered the dataset accordingly. As a result of this extraction process, our analysis targeted 18,681,803 posts by 137,822 active users and 4,103,692 following edges. Table \ref{tab:emotional_cat_summary} presents the seven target emotional categories, including their representative labels and the distribution of posts (number and percentage) among active users.

\section{Analysis of Timeline Emotional Associations}\label{sec:contagion}



This section examines temporal associations between emotions in users' pre-posting timelines and their subsequent emotional labels. We analyze both within-category associations, where the same emotion appears in the timeline and the subsequent post, and cross-category associations, where different emotion categories are involved. We focus on seven emotional categories: Sadness, Fear, Anger, Surprise, Positivity, Affection, and Happiness.

\subsection{Quantification of Pre-posting Emotional Exposure}

First, we define the timeline immediately preceding user $u$'s post $t$ as the set of posts $H_t$ made by $u$'s followees within three hours prior to $t$. To mitigate noise from timelines with extremely sparse activity, we only included posts $t$ where the number of posts in the timeline $|H_t|$ exceeded 20. The final analysis targeted a total of 1,035,304 posts from 29,454 users.

Consistent with existing research on the contagion of positive and negative affects \cite{ferrara2015}, we calculated a baseline distribution representing the average emotional distribution on timelines and determined the difference between the actual distribution in $H_t$ and this baseline. The baseline distribution was calculated as follows: First, all $H_t$ sets associated with all analyzed posts were aggregated into a single population. For each post $t$ by user $u$, a virtual timeline was constructed by randomly sampling a number of posts from this population equal to the actual timeline size $|H_t|$. The average emotional distribution across these virtual timelines was defined as the baseline distribution. We confirmed that the standard error for all emotional categories during this calculation was less than 0.0001, ensuring sufficient statistical precision for subsequent analyses.


\subsection{Within- and Cross-Category Emotional Associations}

\begin{figure}[tb]
    \centering
    \begin{subfigure}[c]{0.59\linewidth}
        \centering
        \includegraphics[width=\linewidth]{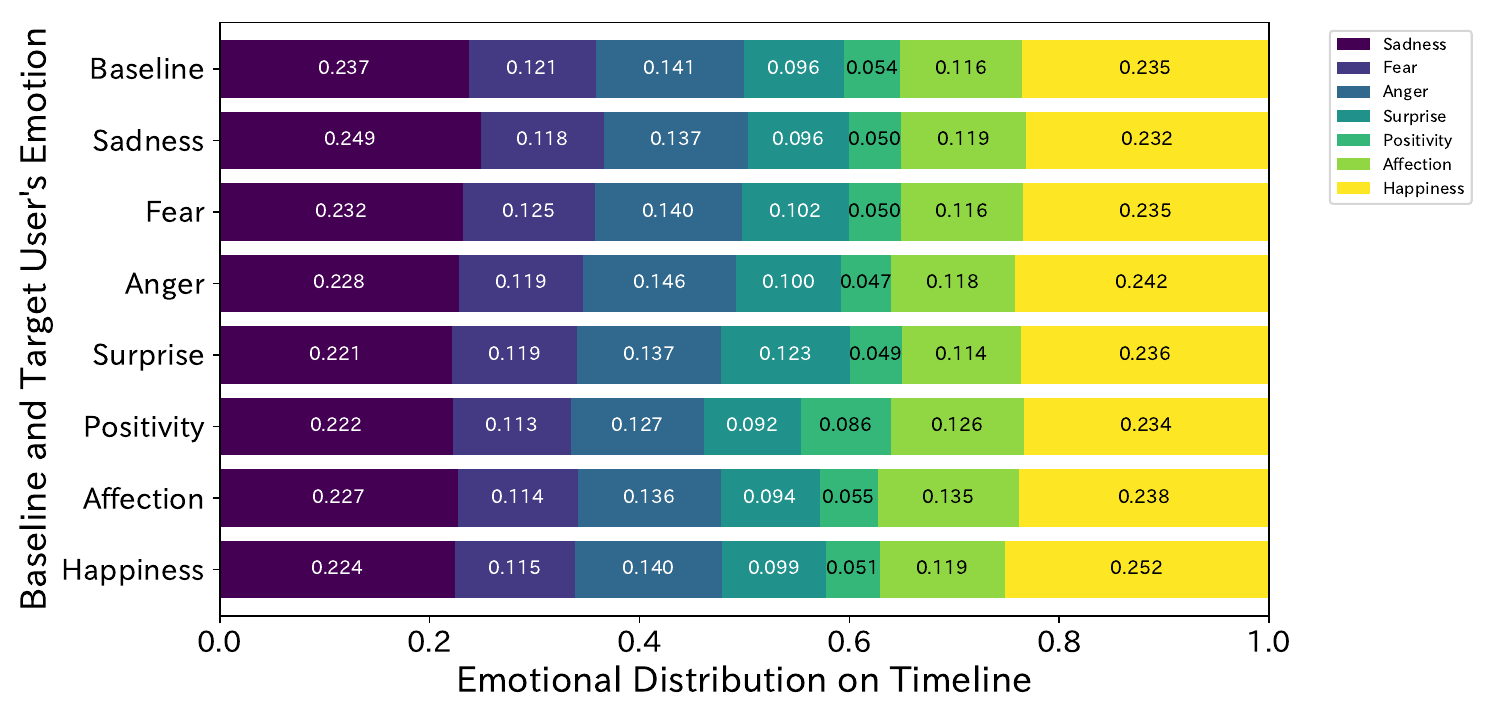}
        \caption{Comparison with baseline}
        \label{fig:contagion_bar}
    \end{subfigure}
    \hfill
    \begin{subfigure}[c]{0.40\linewidth}
        \centering
        \includegraphics[width=\linewidth]{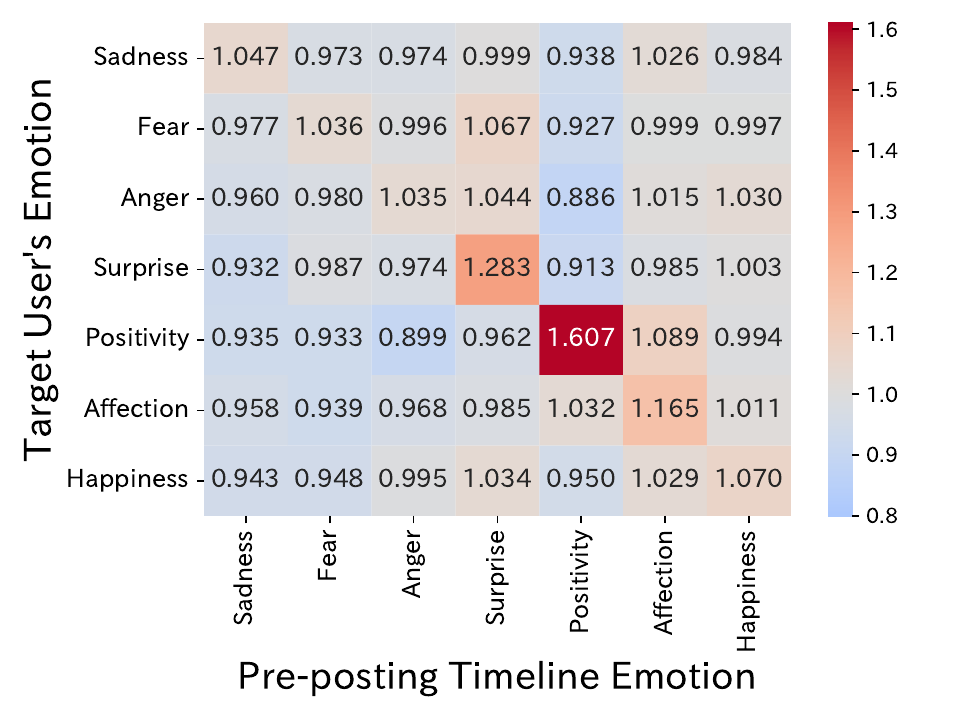}
        \caption{Ratio to baseline}
        \label{fig:contagion_ratio}
    \end{subfigure}
    
    \caption{Emotional distribution in the timeline immediately prior to posting: (a) comparison of absolute distributions and (b) ratio of the timeline distribution to the baseline.}
    \label{fig:contagion_combined}
\end{figure}

To clarify how user emotions are influenced by their preceding timelines, we analyzed emotional biases just before a post is made. Figure \ref{fig:contagion_bar} compares the emotional distribution of pre-posting timelines with the baseline, and Figure \ref{fig:contagion_ratio} illustrates the ratio between these distributions. Mann-Whitney U tests confirmed significant differences from the baseline for all combinations of timeline emotions and user posting emotions.

As shown in Figure \ref{fig:contagion_ratio}, the proportion of the same emotion in the timeline increases prior to a post for all categories, suggesting the occurrence of intra-emotional contagion on Vent. Notably, the proportion of Positivity in the timeline showed the largest relative increase—1.607 times the baseline—just before a user posted a Positivity label. This indicates that Positivity possesses the highest infectivity among the seven analyzed emotions, aligning with prior findings that positive emotions spread more readily than negative ones \cite{ferrara2015,coviello2014}. The second highest increase was observed for Surprise (1.283 times). Berger et al. \cite{berger2012} noted that the virality of online content depends not only on valence (positive vs. negative) but also on arousal—the degree to which an emotion activates and drives an individual to action. This suggests that the high-arousal nature of Surprise contributes to its high transmissibility.

Furthermore, Figure \ref{fig:contagion_ratio} suggests that users' subsequent emotional labels are associated not only with the same emotion but also with different emotion categories in their pre-posting timelines. First, before Fear posts, the proportion of Surprise in the timeline increased by a factor of 1.067. This pattern is consistent with prior work suggesting that surprise can be related to uncertainty and may precede other emotional responses in social media discussions \cite{xing2025,smith1985}. Second, before Anger posts, Affection and Happiness increased by factors of 1.015 and 1.030, respectively. This may be related to social comparison, as prior work has reported that passively viewing others' happiness on Facebook can be associated with envy \cite{krasnova2013}. Third, before Sadness posts, the proportion of Affection was 1.026 times higher. Since the Affection category includes labels related to longing and bittersweetness, such as Needy, this association may reflect emotional proximity between affectionate and sad expressions.

Collectively, these findings answer RQ1 by showing that pre-posting timeline emotions are associated with users' subsequent emotional labels both within and across emotion categories. These cross-category patterns should be interpreted as observational associations rather than causal mechanisms.

\section{Analysis of Susceptibility}

The analysis results in Section \ref{sec:contagion} suggest that, as an average trend across all users, the emotional distribution on the timeline may influence the emotions expressed in their posts. However, it is expected that susceptibility to such influence varies among individual users. To address RQ2, this section investigates the differences in user susceptibility to emotions on their timelines and explores how this susceptibility is characterized.

\subsection{Quantification and Distribution of Susceptibility}

Following Ferrara et al. \cite{ferrara2015}, we quantified users' susceptibility to emotional contagion by examining whether their expressed emotions aligned with recent emotional fluctuations in their timelines. For each post by a user, we first computed the emotional distribution of posts appearing in the user's timeline during the preceding three hours and compared it with the baseline emotional distribution. We then identified the emotion that showed the largest increase relative to the baseline. A match was defined as a case in which the emotion expressed in the user's post was identical to this most strongly increased emotion in the preceding timeline. The susceptibility index is then derived from the frequency of these matches across all of a user's posts. A higher value of this index indicates that the user's emotional expression is more likely to align with short-term emotional fluctuations in their surrounding information environment, suggesting greater susceptibility to emotional contagion. To eliminate the influence of users with a low volume of activity, the final analysis targets 6,908 users who have made at least 20 qualifying posts ($t$) where the number of preceding timeline posts within three hours ($|H_t|$) exceeded 20, consistent with the criteria established in Section \ref{sec:contagion}.

\begin{figure}[tb]
    \centering
    \begin{subfigure}{0.48\textwidth}
        \centering
        \includegraphics[width=\textwidth]{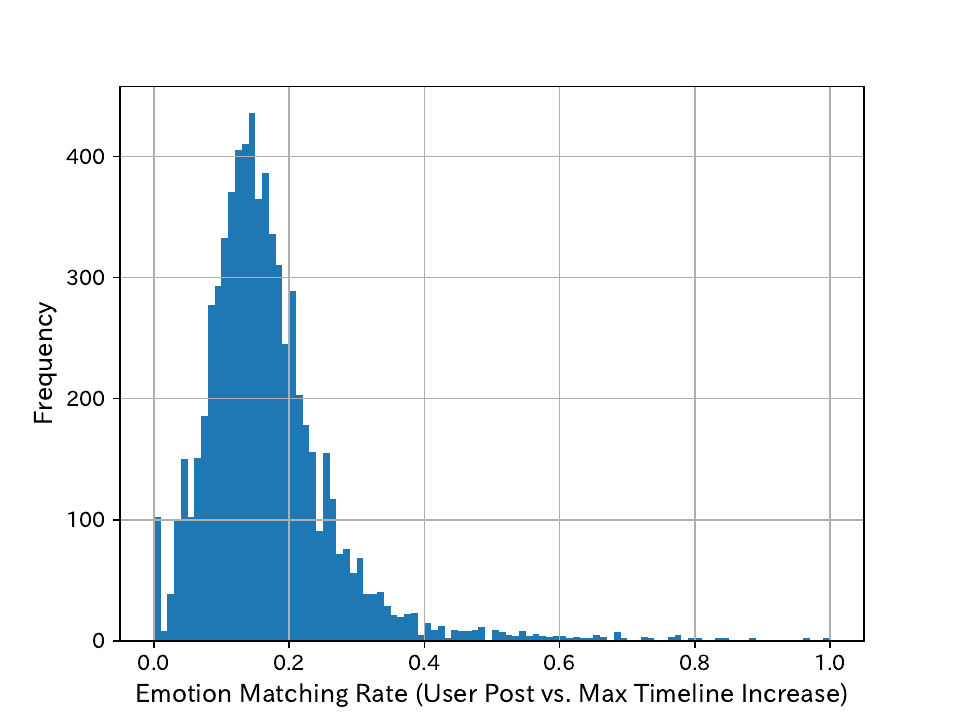}
        \caption{Histogram}
    \end{subfigure}
    \hfill
    \begin{subfigure}{0.48\textwidth}
        \centering
        \includegraphics[width=\textwidth]{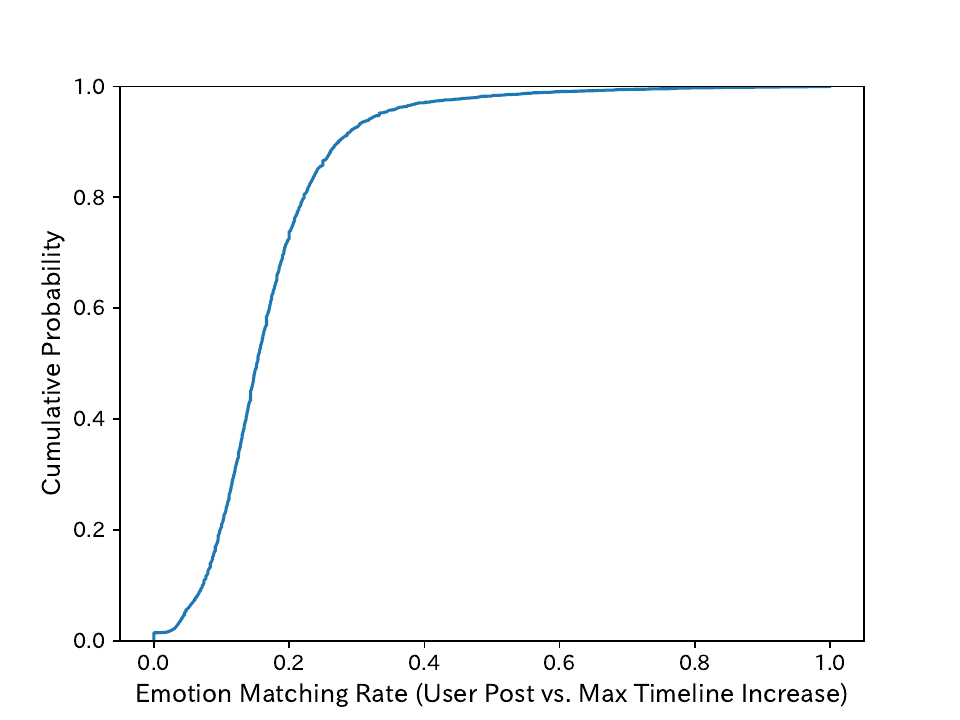}
        \caption{Cumulative Distribution}
    \end{subfigure}
    \caption{Distribution of matching rates between the posted emotion and the emotion with the highest increase rate on the timeline. The distribution is right-skewed, with a small number of users exhibiting high matching rates.}
    \label{fig:sensitivity}
\end{figure}

Figure \ref{fig:sensitivity} presents the histogram and cumulative distribution of the match rate between the emotion posted by a user and the emotion with the highest increase rate on their timeline. From Figure \ref{fig:sensitivity}, it is evident that a small number of highly susceptible users exist who react sensitively to timeline trends and exhibit extremely high match rates. This user segment is at risk of excessively synchronizing with the negative emotional expressions of their surroundings, which may result in increased psychological distress. 


\subsection{Homophily in Susceptibility}\label{sec:high_distance}

It is known that homophily exists in susceptibility to information diffusion—referring to the degree to which a user reacts to information from their surroundings—and that individuals with similar levels of susceptibility tend to connect with one another \cite{luceri2025,tamura2026}. In this section, we examine whether a similar tendency can be observed for users' alignment with emotional fluctuations in their timelines. For the following network analysis, we define high-susceptibility users as those in the top 15\% of the susceptibility index.

First, we calculated the correlation between each user's susceptibility and that of their followees. The correlation coefficient was 0.437, confirming a moderate positive correlation. This finding suggests that users with similar levels of susceptibility to emotional contagion tend to be connected within the network.


Next, to clarify the proximity between highly susceptible users, we calculated the distribution of shortest path lengths for all pairs of high-susceptibility users. As a baseline for comparison, we constructed a null model by randomly rewiring edges connected to high-susceptibility users while preserving their original degrees. The values for the null model represent the average of 10 trials. The results are presented in Tab. \ref{tab:high_hops}. The number of high-susceptibility user pairs in a direct following relationship (1-hop) was approximately 8.3 times higher than that in the null model. Furthermore, the average number of hops between high-susceptibility users was 2.35, shorter than the 3.29 hops observed in the null model. Additionally, while only about 63.0\% of pairs were located within three hops in the null model, this proportion reached approximately 99.7\% among high-susceptibility users.


These results indicate that high-susceptibility users are more strongly clustered in the network than expected under the null model. Such proximity may create conditions for local reinforcement of emotional expressions, although the present analysis does not directly establish an echo-chamber effect.

\begin{table}[tb]
\centering
\caption{Number of high-susceptibility user pairs by shortest path length compared to the null model}
\label{tab:high_hops}
\begin{tabular}{lrr}
\toprule
       & High Susceptibility & Null Model (SEM) \\
       \midrule
    1 hop & 3,782 & 457.4 (5.6) \\
    2 hops & 345,503 & 45,227.1 (87.5) \\
    3 hops & 188,547 & 293,805.9 (1708.8) \\
    4 hops & 1,406 & 199,747.4 (1754.2) \\
    5 hops & 3 & 3.2 (2.0) \\
    \bottomrule
  \end{tabular}
\end{table}
  \hfill

\section{Conclusion}
In this paper, we analyzed emotional dynamics on Vent, an emotion-sharing social media platform with explicit user-assigned emotion labels. For RQ1, we found that users' subsequent emotional labels are associated with the emotional composition of their pre-posting timelines, with both within-category and cross-category associations observed among several emotion categories. For RQ2, we quantified users' alignment with timeline emotional fluctuations and found that highly aligned users tend to be located close to one another in the network compared with a degree-preserving null model. Overall, these findings provide large-scale observational evidence that emotional expression on Vent is related to both short-term timeline context and network structure. Because our analysis is based on followee posts within a preceding time window, it captures potential rather than actual exposure. Moreover, the observed associations may also reflect stable user-specific emotion profiles, emotional homophily, or common external events, rather than short-term emotional contagion alone. Future work should further examine causal identification and the generalizability of these patterns to other social media platforms.

\begin{credits}
\subsubsection{\ackname}
This work was supported by JSPS KAKENHI Grant Number JP25K03105 and JST ERATO Grant Number JPMJER2502.

\subsubsection{\discintname}
The authors have no competing interests to declare that are relevant to the content of this article.
\end{credits}

\bibliographystyle{splncs04}
\bibliography{references}

\end{document}